\documentclass[prl,onecolumn,aps,10pt,notitlepage]{revtex4-1}

\usepackage{amsmath,amssymb}
\usepackage{graphicx}
\usepackage{slashed}
\usepackage{subfigure}

\begin{document}

\title{Direct numerical approach to one-loop amplitudes}
\author{G.~Duplan\v ci\'c}
\email{Goran.Duplancic@irb.hr}
\author{B.~Klajn}
\email{Bruno.Klajn@irb.hr}
\affiliation{Theoretical Physics Division, Rudjer Boskovic Institute, Bijenicka cesta 54, HR-10000 Zagreb, Croatia}

\date{\today}

\begin{abstract}
We present a completely numerical method of calculating one-loop amplitudes. Our approach is built upon two different existing methods: the contour deformation and the extrapolation methods. Taking the best features of each of them, we devise an intuitive, stable and robust procedure which circumvents the problem of large cancellations and related numerical instabilities by calculating the complete amplitude at once. As a proof of concept, we use our method to calculate the $2\gamma \to (N - 2)\gamma$ benchmark process, as well as the Higgs decay amplitude $H \to \gamma \gamma$.
\end{abstract}

\maketitle

\section{Introduction}

In the framework of quantum field theory, obtaining radiative corrections requires the evaluation of loop Feynman amplitudes. 
The simplest loop amplitudes, but also the most important ones, are the one-loop Feynman amplitudes which are usually associated with the
evaluation of the next-to-leading-order virtual corrections.

The traditional way of evaluating the one-loop Feynman amplitudes follows a familiar scheme. 
First the amplitude contributions are determined by writing down all relevant Feynman diagrams. By using Feynman rules, each diagram
is then translated into an explicit mathematical expression in the form of a Feynman integral. 
In the next step, using various recursion relations and procedures, e.g.~the widely known Passarino-Veltman reduction, 
all, i.e.~tensor and scalar, one-loop Feynman integrals are decomposed down to a linear combination of an independent set of 
master integrals. Usually, the basic set of master integrals is comprised of one-loop scalar Feynman integrals with up to four lines.
These master integrals have been well examined and their analytical expressions and properties are well known. 
As the last step, the expressions for the contributing Feynman diagrams obtained this way are summed to give the final result for 
the amplitude. 

Thanks to its universality and complete mathematical understanding, the traditional approach, i.e.~representing the amplitude 
in terms of the basic master integrals, became the paradigm for evaluation of one-loop amplitudes. 
However, decomposition of particular diagrams (and consequently the complete amplitude) down to 
the master integrals can be very demanding, even if computers are used for the task. 
Hence, development of an efficient technique for evaluation of the coefficients 
in the decomposition is in the focus of interest for all practitioners in the field, 
especially due to the fact that the precision and quality of the recent experimental data are so high that complete and trustworthy comparison 
with the theoretical predictions requires at least next-to-leading order calculations for a large variety of processes.
Over the last two decades we have witnessed intense development of efficient techniques taking care of stability, speed and precision.
Recently popular techniques, publicly available as computer codes and tested in many physical scattering processes, 
are based either on integration-by-parts relations or unitarity \cite{helac,mg5,gosam1,gosam2,openloops,recola}. 

Since the final representation of the diagram or the amplitude 
as a linear combination of an independent set of master integrals is unique, all traditionally oriented techniques have the same 
shortcomings despite the particular technique or implementation (analytical or numerical) used to determine the coefficients.
For large number ($N > 4$) of external particles, the complexity and number of contributing diagrams grows rapidly. Consequently, the number
of master integrals needed for the diagram and amplitude decomposition also increases quickly. All this makes 
the calculation very demanding and repetitive, particularly in the case where the calculation is performed diagram by diagram. 
Namely, it is known that the final expression for the coefficients in the
amplitude is, more often than not, much simpler than the bare sum of the expressions of the contributing diagrams themselves. 
This means that there are large cancellations between the contributions of different diagrams, 
implying that a lot of laborious calculation is probably avoidable.
Furthermore, during reduction to master scalar integrals, many Gram and modified Cayley determinants of various powers
are generated in the denominators of the decomposition coefficients. Vanishing of these determinants, particularly Gram determinants,
does not correspond always to real singularities of the diagrams or the complete amplitude. Consequently, 
in a kinematical region where some of the determinants are small, one faces cancellation of big numbers and numerical instabilities.

Taking all that into account, it is not clear whether the reduction to master integrals is the optimal method
of performing the calculations, especially because, due to their complexity, the calculations become more and more 
numerically oriented, while the traditional approach is  
essentially analytically oriented and designed to avoid direct integration for as long as possible.
It would seem that some form of a direct numerical integration of the Feynman integrals is a more natural approach for the
numerical calculations and, moreover, it may also be more efficient. At the very least, the efficiency may stem from the fact that the calculation of 
the physical cross sections requires a direct numerical integration of the squared amplitudes over the momenta of the final state particles.
This integration can be combined with the integration of the Feynman integrals and performed simultaneously.   
Still, there are some difficulties related to the numerical approach.
Physical amplitudes can exhibit both UV and IR divergences that have to be isolated before numerical integration. 
Subtraction methods \cite{soper3, weinzriel3, weinzriel5} provide a general approach to handle this kind of problem. 
Another difficulty is related to stability and convergence of the numerical integration. Namely, getting a sufficiently precise result using a 
reasonable number of integrand
evaluations is hard to accomplish due to the poles and cuts on the integration path, cancellations
between integrands, high-dimensional integration domains and integrable threshold singularities. 
Recently, the contour deformation method \cite{soper2, soper1, weinzriel3, weinzriel1, weinzriel2, weinzriel5} has been devised to deal with 
the issues mentioned above. However, the deformation procedure is involved, with a lot of
peculiarities and details to be taken into account in order to deal with all cases of practical interest at an acceptable level. 
As a rule, the contour deformations are simpler if the amplitude is represented as an integral over Feynman parameters but the convergence is better
if the integration is done directly in the momentum space, where the contour deformations are much more complicated. Also, deformations become
more cumbersome whenever massive particles propagate in the loop.
On the other hand, the extrapolation method \cite{doncker6, doncker5, doncker2} is more intuitive and straightforward, but 
the way it has been implemented, using Feynman parametrization in diagram-by-diagram approach, lacks the possibility of an effective 
calculation of the complete amplitude.

All things considered, it is clear that both approaches (reduction to master integrals and direct numerical integration) have some intrinsic
disadvantages originating from unavoidable singular behavior of the Feynman amplitudes. However, relying on insights from extrapolation
method, we still believe it is     
justified to try to formulate a direct numerical
integration method which should be intuitive, easy to implement, generally applicable, capable of handling complete amplitude at once and which
gives sufficiently precise results. In this paper we present such a method.

%\medskip

\section{Method}
All one-loop amplitudes are given by a generic expression
\begin{equation}
{\cal M} = \sum_{i} \int \frac{{\rm d}^4 \ell}{(2 {\pi})^4} 
\frac{{\cal N}_i(\ell_\mu,\,\ldots)}{
\prod_{j=1}^{n_i} \left( \left( \ell+r_{i j}\right) ^2-m^2_{i j}+{\rm i}\epsilon \right)},\label{1}
\end{equation}
where $i$ denotes contributions arising from different one-loop diagrams, ${\cal N}_i(\ell_\mu,\,\ldots)$ is the polynomial function of the loop-momentum components $\ell_\mu$ over which we integrate, $n_i$ is the number of internal lines in the $i$th diagram loop, $r_{ij}$ are the linear combinations of the momenta of the external particles, $m_{ij}$ are the masses of the particles going through the loop, and infinitesimal imaginary part ${\rm i}\epsilon$ (with $\epsilon >0$) 
is the so-called Feynman prescription which ensures causality. Physical amplitudes are defined in the limit $\epsilon \to 0$.

The method is based on the following reasoning. It is known that there are cancellations between different 
contributions, i.e.~diagrams. In the usual diagram-by-diagram, approach as given by Eq.~(\ref{1}), this requires
a numerical integration of high precision for each diagram to avoid possibly large subtraction-induced errors in the final summation. To this end, it is numerically more efficient to exchange the order of the sum and the integral in Eq.~(\ref{1}). As a consequence, the unnecessary wasting of resources on precise numerical integration could be avoided. Also, it is sensible to integrate directly over the loop momentum without invoking Feynman parameters because the number of
integrations in the Feynman parameter representation is equal to the number of the internal lines in the loop, which is generally not the same for all one-loop contributing diagrams. Additionally, since it is usually the case that $n_i\geq 4$, integrating directly in four dimensions keeps the integration domain low dimensional and, in effect, makes the numerical integration more stable and convergent. However, numerical stability is guaranteed only if the integrand is a smooth function, with no sharp peaks; this is not the case here, see Eq.~(\ref{1}), as the integrand has singularities along the surfaces 
$\left( \ell+r_{i j}\right) ^2=m^2_{i j}$, assuming $\epsilon \to 0$. 

\begin{figure}[!t]
\centering
\includegraphics[scale=1]{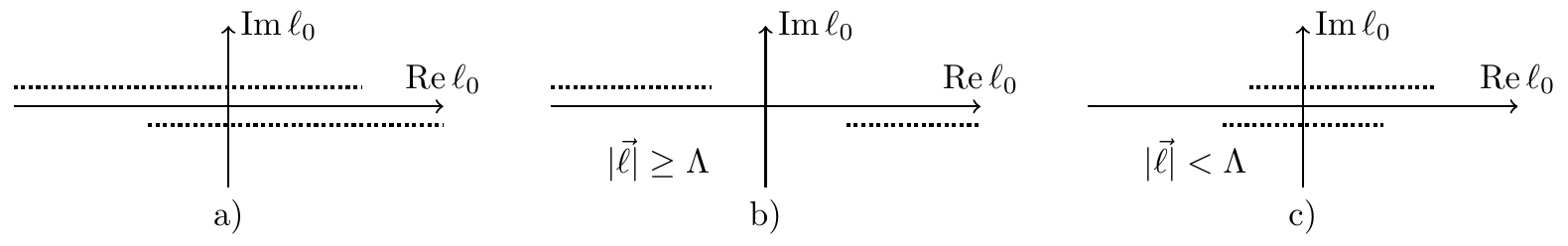}
\caption{\label{fig:1}Schematic position of the integrand poles (\ref{2}) for the amplitude (\ref{1}) in the complex $\ell_0$ plane. The $\Lambda$ is some scale greater than ${\rm max}\,|r_{i j,0}|$. (a) general position of the poles; (b) UV region; (c) IR region.}
\end{figure}

In the contour deformation approach these singularities are avoided by performing deformations of the integration contour in the complex loop momentum $\ell$ space. Here we are aiming for a much simpler approach. First, we also avoid singularities by a contour deformation where it is easy and straightforward to do so. In order to find such regions, we have to examine the singularities in more detail. If the space components of the loop momenta $\vec{\ell}$ are fixed, the singularities of the integrand are poles located at
\begin{equation}
\ell_{0,ij,\pm} = - r_{i j,0}\pm \left[ (\vec{\ell}+\vec{r}_{i j})^2+m_{i j}^2-{\rm i}\epsilon \right]^{1/2}. \label{2}
\end{equation}
%as is depicted in
Figure \ref{fig:1}(a) shows the position of the singularities for the general case (arbitrary $\vec{\ell}$, $r_{ij}$ and $m_{ij}$) in the complex $\ell_0$ plane. It is seen that there is a region where the integration contour can be pinched between the poles making the contour deformation impossible and stable numerical integration difficult, but there is also a region where the singularities can be easily avoided by contour deformation. It is straightforward to determine in what way are these two regions separated. The pinching does not appear as long as $\ell_{0,ij,\pm} \gtrless 0$. By using triangle inequality in Eq.~(\ref{2}), this condition is equivalent to $( |{\vec \ell}|+|\vec{r}_{ij}|)^2 >|r_{i j,\,0}|^2-m_{i j}^2$. It follows that for large enough $|{\vec \ell}|$ (so that $|{\vec \ell}|>{\rm max}\,|r_{i j,0}|$) it is impossible for the integration path to be pinched (see Fig.~\ref{fig:1}.b). On the other hand, for smaller values of $|{\vec \ell}|$, the singularities are localized and placed on both sides of the integration path (see Fig.~\ref{fig:1}.c). Taking all this into account, it is justified to introduce a scale $\Lambda$ to separate these two regions of different properties. The two regions $|{\vec \ell}|\ge\Lambda$ and $|{\vec \ell}|<\Lambda$ are referred to as the UV and the IR regions, respectively.

In the UV region [see Fig.~\ref{fig:1}(b)], the usual Wick rotation is sufficient to move the integration curve away from the poles into the region
where the integrand is a smooth function and a stable numerical integration can be performed. If the amplitude is UV divergent, however, it is necessary to use some subtraction procedure to remove the parts contributing to the divergence from the integrand before numerical integration.

For the IR region [see Fig.~\ref{fig:1}(c)] there is no easy procedure to accomplish stable integration. Nevertheless, we choose to adjust the extrapolation method from \cite{doncker6}. Before giving arguments for such a choice, let us explain the method. The extrapolation method relies on the fact that the Feynman integrals $I$ are actually analytical functions of the parameter $\epsilon$. For example, for finite $\epsilon$, integrands of the Eq.~(\ref{1}) do not have singularities in the integration domain, and
become much smoother with lower peaks, which makes the integration more stable and convergent. Therefore, the idea is to calculate a sequence of $I(\epsilon_i)$, $i=1,2,\ldots$, and extrapolate to the limit $I(0)$. Originally, this method is used to evaluate particular Feynman diagrams, i.e.~integrals in the Feynman parameters representation and it is shown \cite{doncker6} that the procedure gives good predictions for relatively large values of $\epsilon_i$, which is exactly what we need. Namely, we are trying to evaluate, i.e.~integrate, the complete amplitude at once. We can count on the lack of large cancellations between diagrams, contrary to the diagram-by-diagram approach, but our integrand is much bigger. In regards to that, it is more important to ensure stability than high precision in the integration. Knowing that large $\epsilon_i$ stabilize the integration, and relying on the above-mentioned characteristics of the extrapolating procedure, there is a real chance of getting a sufficiently precise result with a reasonable number of integrand evaluations if a simple and basic extrapolation method is used. The only possible unknown is the choice of the sequence $\{\epsilon_i\}$ for which the IR part of the amplitude has to be calculated in order to achieve a correct limit for $\epsilon \to 0$.

To summarize before proceeding with the analysis, in order to have a stable numerical evaluation, we expressed the amplitude (\ref{1}) in the form 
\begin{equation}
{\cal M} =  {\rm i}\int\limits_{-\infty}^{+\infty} {\rm d} \ell_0 \int\limits_{|{\vec \ell}|\ge\Lambda} {\rm d^3} \ell\,\sum_{i}{\cal I}_i\left({\rm i}\ell_0,
\vec{\ell},\epsilon=0\right) +  \lim_{\epsilon\to 0}\, \int\limits_{-\infty}^{+\infty} {\rm d} \ell_0 \int\limits_{|{\vec \ell}|<\Lambda} {\rm d^3} \ell\,\sum_{i}{\cal I}_i\left(\ell_0,\vec{\ell},\epsilon\right), \label{3}
\end{equation}
where ${\cal I}_i$ stand for the integrands of the Eq.~(\ref{1}). The UV part is straightforward to integrate and for the IR part we need a fast and simple integration method because the integral has to be evaluated for the set of different $\epsilon_i$. From several considered methods it appears that the adaptive Monte Carlo method is the best approach. Also, although the IR part can be integrated directly as given in Eq.~(\ref{3}), it is advantageous to perform the $\ell_0$ integration using the residue theorem. The remaining integral is three dimensional and has a finite integration domain which consequently means a fewer number of Monte Carlo evaluation points for the desired precision. However, there are some subtleties to this. We assume all the poles of the integrand $\sum_{i}{\cal I}_i\left(\ell_0,\vec{\ell},\epsilon\right)$ at points (\ref{2}) are of the first order which is not generally correct. However, since evaluation points in the Monte Carlo integration are chosen at random, there is a small probability of hitting the pole of higher order exactly. Moreover, hitting a point near such a pole numerically approximates the derivative of the first-order residue, which, indeed, is the correct form of the higher-order residue. Finally, the amplitude prepared for the numerical integration is given in the form
\begin{equation}
{\cal M} =  {\rm i}\int\limits_{-\infty}^{+\infty} {\rm d} \ell_0 \int\limits_{|{\vec \ell}|\ge\Lambda} {\rm d^3} \ell\,\sum_{i}{\cal I}_i\left({\rm i}\ell_0,
\vec{\ell},\epsilon=0\right)+ \lim_{\epsilon\to 0}\, \int\limits_{|{\vec \ell}|<\Lambda} {\rm d^3} \ell\, 2\pi {\rm i}\sum {\rm Res} \left(\sum_{i}
{\cal I}_i\left(\ell_0,
\vec{\ell},\epsilon\right), \ell_0= \ell_{0,ij,-} \right). \label{4}
\end{equation}

The last thing to be discussed is how to perform the extrapolation to get $\lim_{\epsilon\to 0}\,{\cal M}_{\rm IR}(\epsilon)$ where ${\cal M}_{\rm IR}(\epsilon)$ stands for the integral in the second term of the Eq.~(\ref{4}).
Originally in the extrapolation method \cite{doncker6} Wynn's epsilon algorithm is used. However, for our approach where complete amplitude is integrated at once, it is hard to predict how fast the amplitude ${\cal M}_{\rm IR}(\epsilon)$ will converge for $\epsilon \to 0$, for different processes and kinematics. As it is possible that amplitude is a steeper function near $\epsilon=0$, in the end we decided to use the Pad\'e approximant (the best approximation of a function by a rational function) as a generally acceptable approach to approximate the ${\cal M}_{\rm IR}(\epsilon)$ and determine ${\cal M}_{\rm IR}(0)$ from numerically evaluated ${\cal M}_{\rm IR}(\epsilon_i)$.

\begin{figure}[!t]
\centering
\includegraphics[scale=1.2]{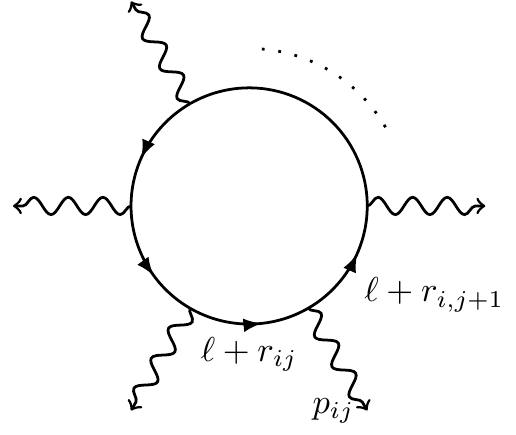}
\caption{\label{loop}A generic Feynman diagram for the $N$-photon amplitude.}
\end{figure}

%\medskip

\section{Examples}

We experimented with different processes and kinematics, varying number of Monte Carlo points and $\{\epsilon_i\}$ sequences, and, to our surprise, the method described above appears quite robust and stable. By that we mean that it is possible to get a very good agreement between our method and analytical results by using relatively small number of integration points and large values of $\epsilon_i$ (details are given below). Here we present it by taking as example the amplitudes for $2\,\gamma\to (N-2)\gamma$ through a electron loop and the Higgs decay amplitude $H \to \gamma \gamma$ through a $W$-boson loop.

\subsection{Photon amplitudes}

The one-loop process $2\,\gamma\to (N-2)\gamma$ is used as a benchmark for testing numerical one-loop amplitudes \cite{soper2, soper1}. It is, therefore, useful as a test of our method where complete amplitude is calculated at once because there are integrable infrared divergences (for the massless case), ultraviolet divergences should cancel (for $N=4$) and the number of contributing diagrams is not small (120 for $N=6$). In the standard notation of Eq.~(\ref{1}), the amplitude for $2\,\gamma\to (N-2)\gamma$ is given as a sum  of the contributions depicted in Fig.~\ref{loop},
\begin{equation}
{\cal M}={\rm i}\sum_{i=1}^{(N-1)!} \int \frac{d^4 \ell}{(2 \pi)^4}e^N {\cal N}_i(\ell) \prod^N_{j=1}\frac{1}{(\ell+r_{ij})^2-m^2+{\rm i}\epsilon},
\end{equation}
where $i$ is going over all permutations of the external lines, ${\cal N}_i(\ell)$ is the numerator function of the form ${\cal N}_i(\ell)={\rm Tr}[\slashed{\varepsilon}_N (\slashed{\ell}+\slashed{r}_{iN}+m)\ldots
\slashed{\varepsilon}_1 (\slashed{\ell}+\slashed{r}_{i1}+m)]$ where the photon polarization $\varepsilon_j$ should be conjugated if the corresponding photon is outgoing.
Here we use the same kinematics and helicities as in \cite{soper2, soper1}.
The momenta of the incoming photons are along the $z$ axis and for the $N=6$
case the final state momenta are chosen as
\begin{eqnarray}
{\vec p_3}=(33.5,\,15.9,\,25),&\quad &
{\vec p_4}=(-12.5,\,15.3,\,0.3)\label{state} \\
{\vec p_5}=(-10,\,-18,\,-3.3),&\quad  &
{\vec p_6}=(-11,\,-13.2,\,-22),\nonumber
\end{eqnarray}
and new momenta configurations are created by rotating 
the final state through the angle $\theta$ about the $y$ axis.

To push the method to the limit we integrate 
the amplitudes ${\cal M}_{\rm IR}(\epsilon_i)$ using just $10^6$ Monte Carlo integration points and asking for precision 
of $10^{-2}$. For comparison, in diagram-by-diagram approach, at least $10^6$ Monte Carlo points and precision higher than $10^{-8}$ is needed for each diagram. It is a reasonable guess that in this case the $\epsilon_i$ should not be smaller than $10^{-4} s$,
where $s$ is the total energy of the process squared. Namely, to obtain reliable results, the peak structure of the integrand should be wide enough to be ``visible'' by the imposed density of the Monte Carlo points. In the center of the mass frame ${\rm max}\,|r_{i j,\,0}|=\sqrt{s}$ 
and, to be on the safe side, here we choose a larger scale $\Lambda=2\sqrt{s}$. For extrapolation we use 
the Pad\'e approximant of order $[ 2/1 ]$ which is a simple rational function 
$R_{(2/1)}(\epsilon)=(a_0+a_1 \epsilon+a_2 \epsilon^2)/(1+b_1 \epsilon)$. This Pad\'e approximant is good enough to reproduce ${\cal M}_{\rm IR}(\epsilon)$ even if it is peaked near $\epsilon = 0$ and, at the same time, it is not too sensitive if ${\cal M}_{\rm IR}(\epsilon_i)$ are calculated with large uncertainties.

\begin{figure}[!h]
\centering
\subfigure{
\includegraphics[width=.35\textwidth]{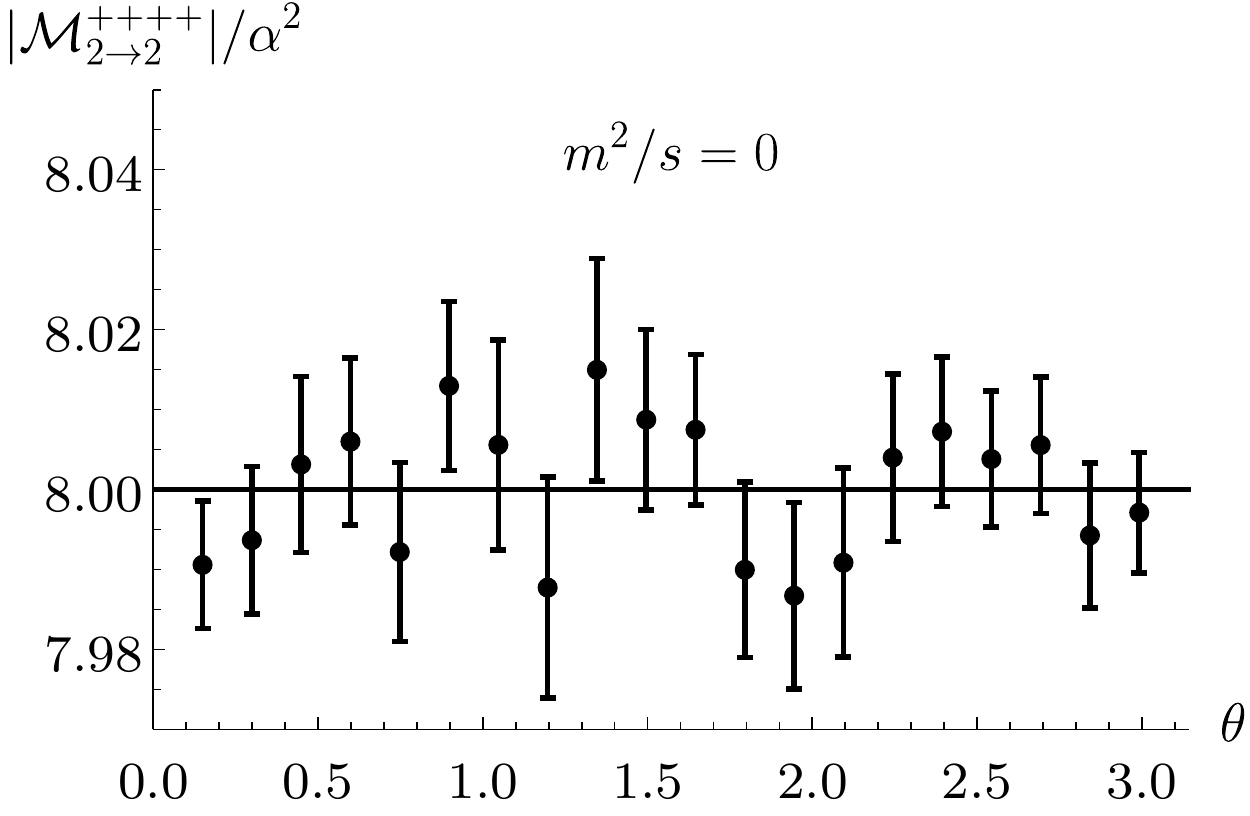}
}
\hspace{50pt}
\subfigure{
\includegraphics[width=.35\textwidth]{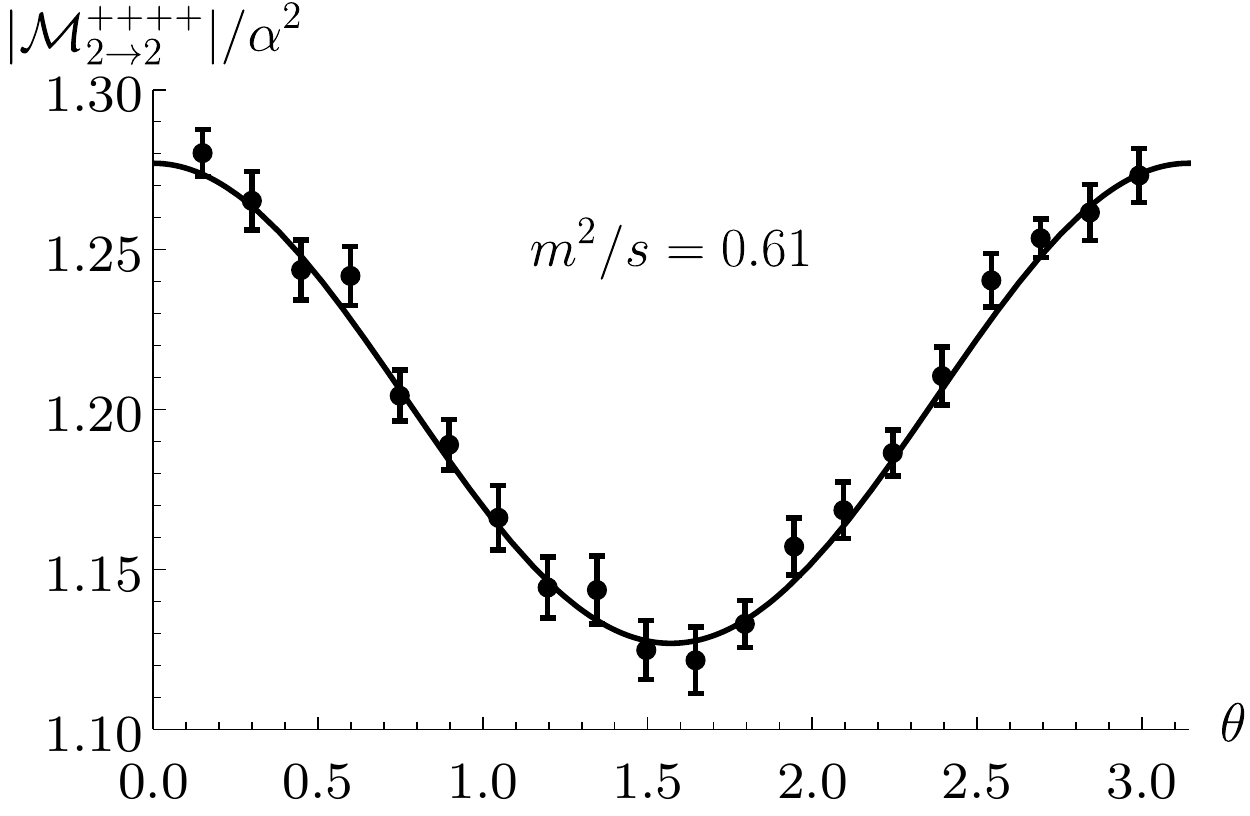}
} \\
\subfigure{
\includegraphics[width=.35\textwidth]{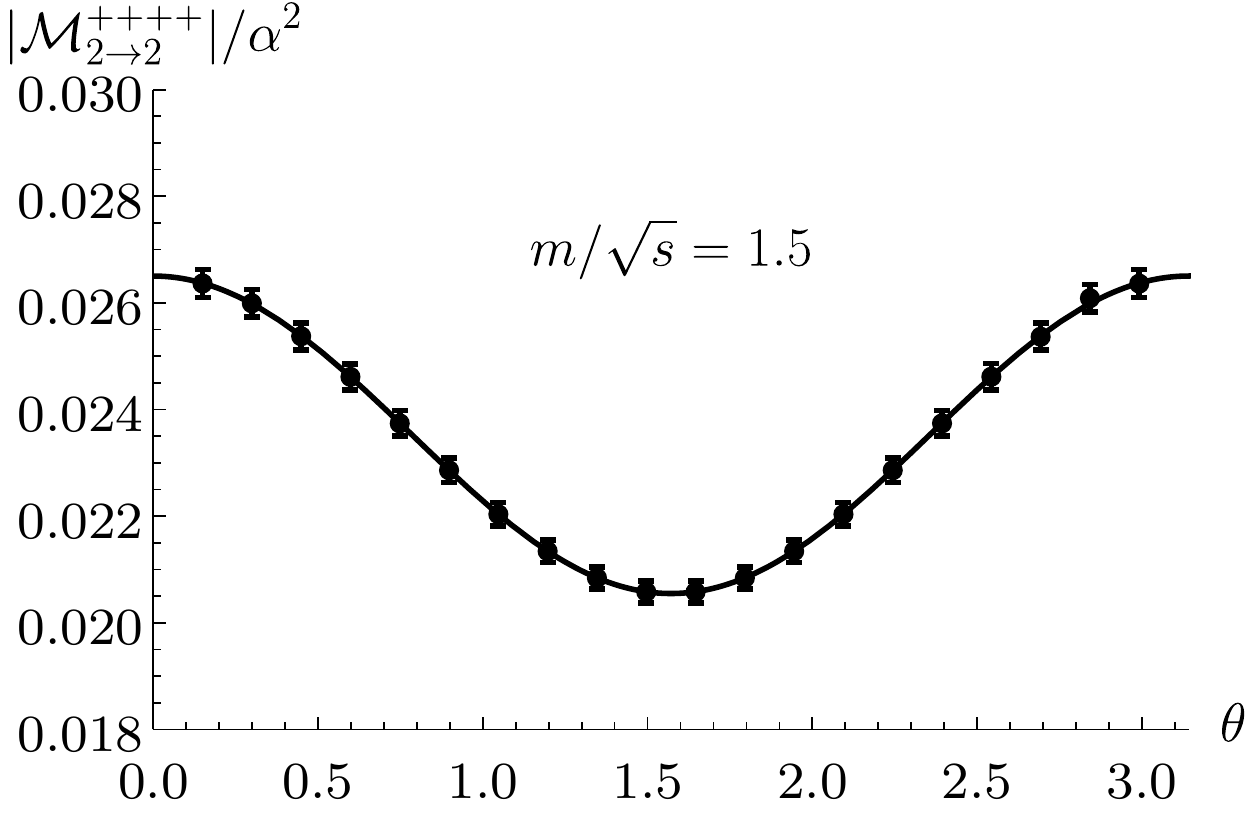}
}
\hspace{50pt}
\subfigure{
\includegraphics[width=.35\textwidth]{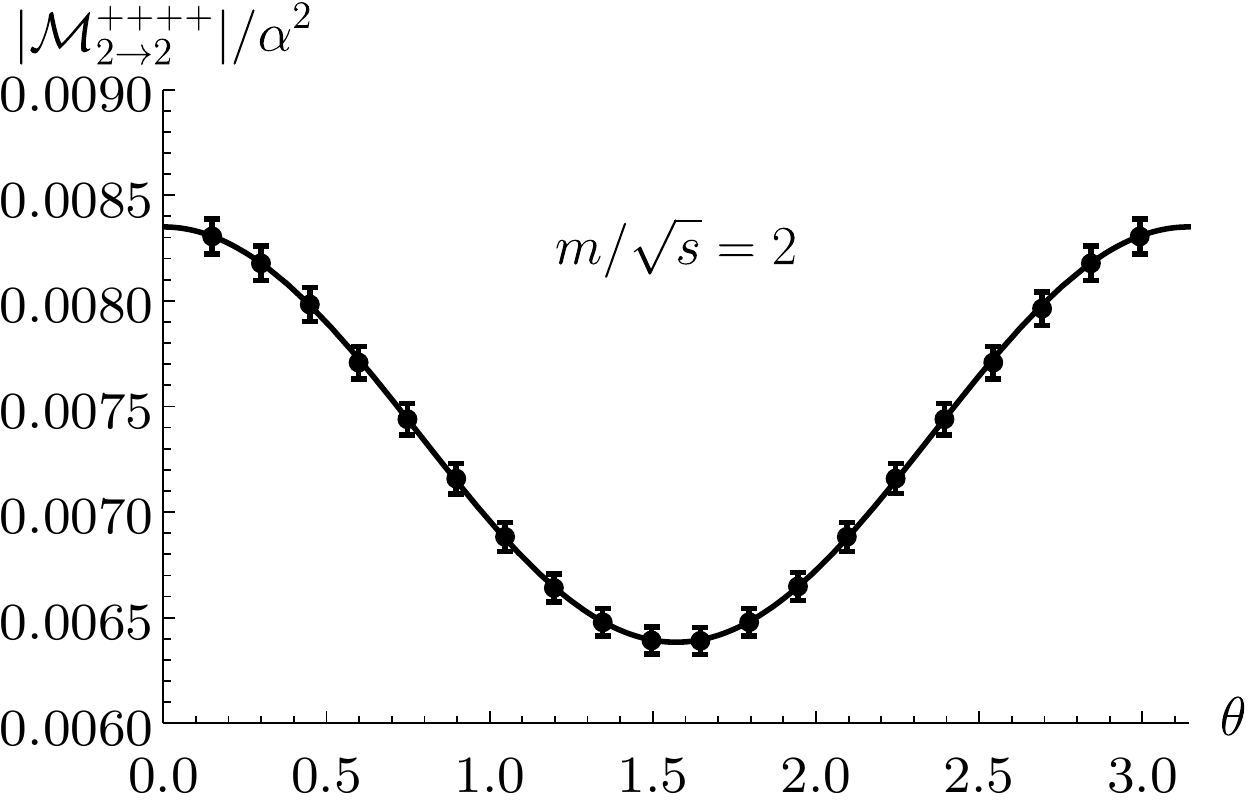}
}
\caption{\label{pppp} Four-photon amplitudes for $++++$ helicities and different fermion masses. 
The curve is the analytic result, 
$\theta$ is the scattering angle
and the points are the result of numerical integration and extrapolation. Note that for $m > \sqrt{s}$, all integrand poles are positioned as in Fig.~\ref{fig:1}(b) and there is no need to separate the integration domain into UV and IR regions. Namely, the complete amplitude can be Wick rotated and directly integrated for $\epsilon = 0$.}
\end{figure}

\begin{figure}[!h]
\centering
\subfigure{
\includegraphics[width=.35\textwidth]{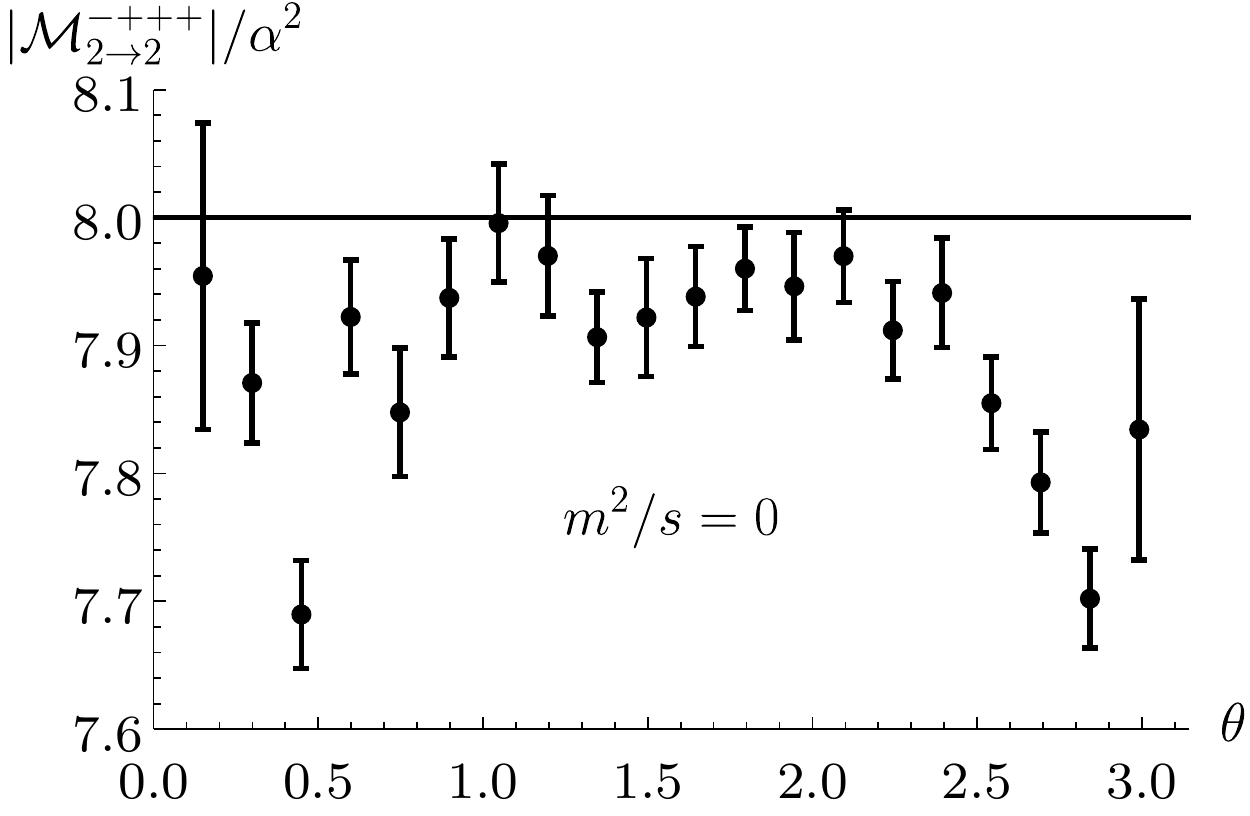}
}
\hspace{50pt}
\subfigure{
\includegraphics[width=.35\textwidth]{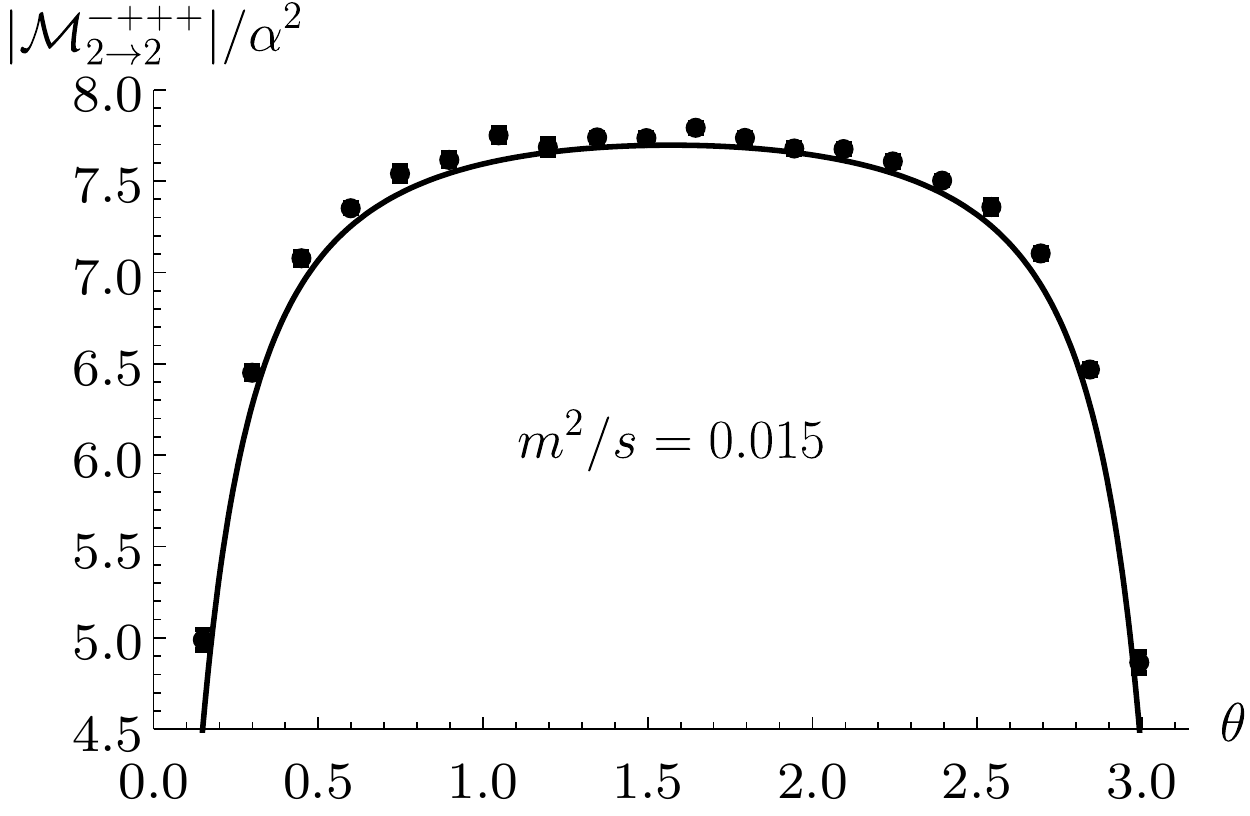}
} \\
\subfigure{
\includegraphics[width=.35\textwidth]{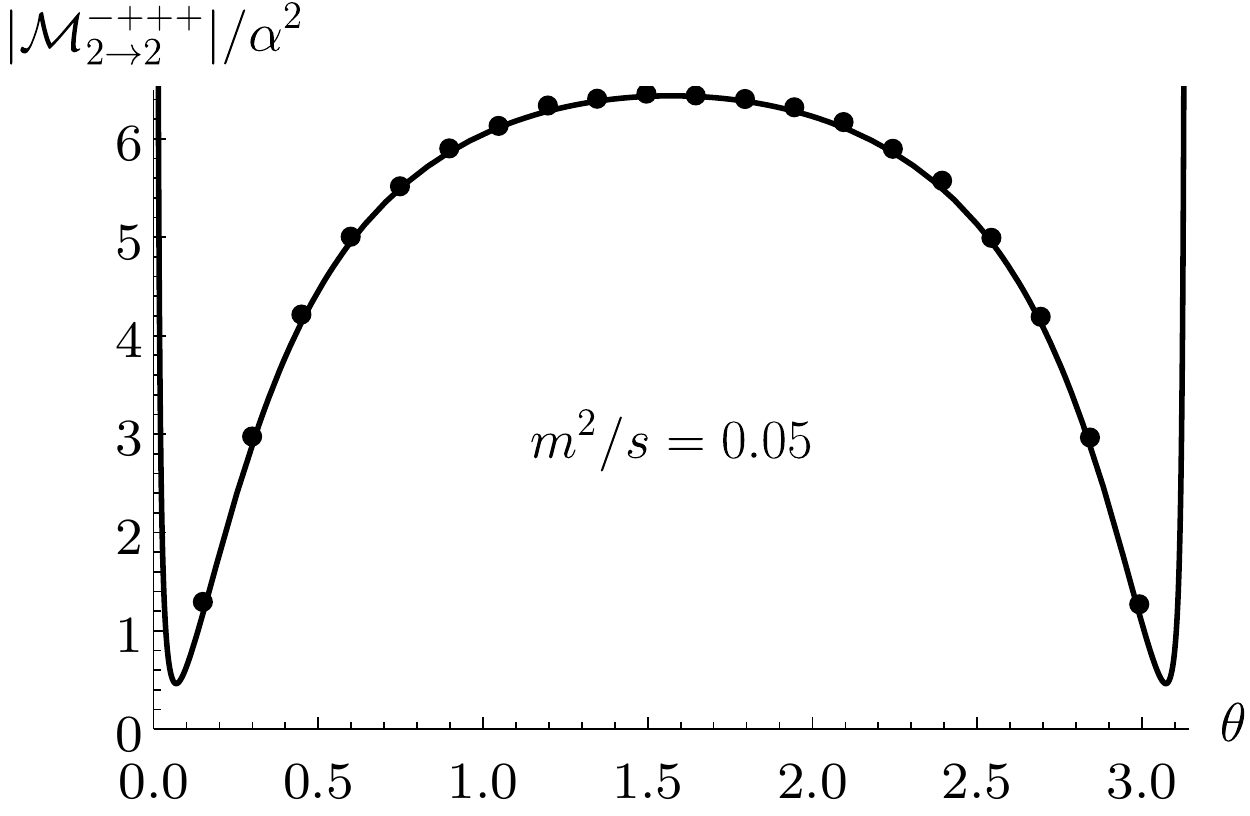}
}
\hspace{50pt}
\subfigure{
\includegraphics[width=.35\textwidth]{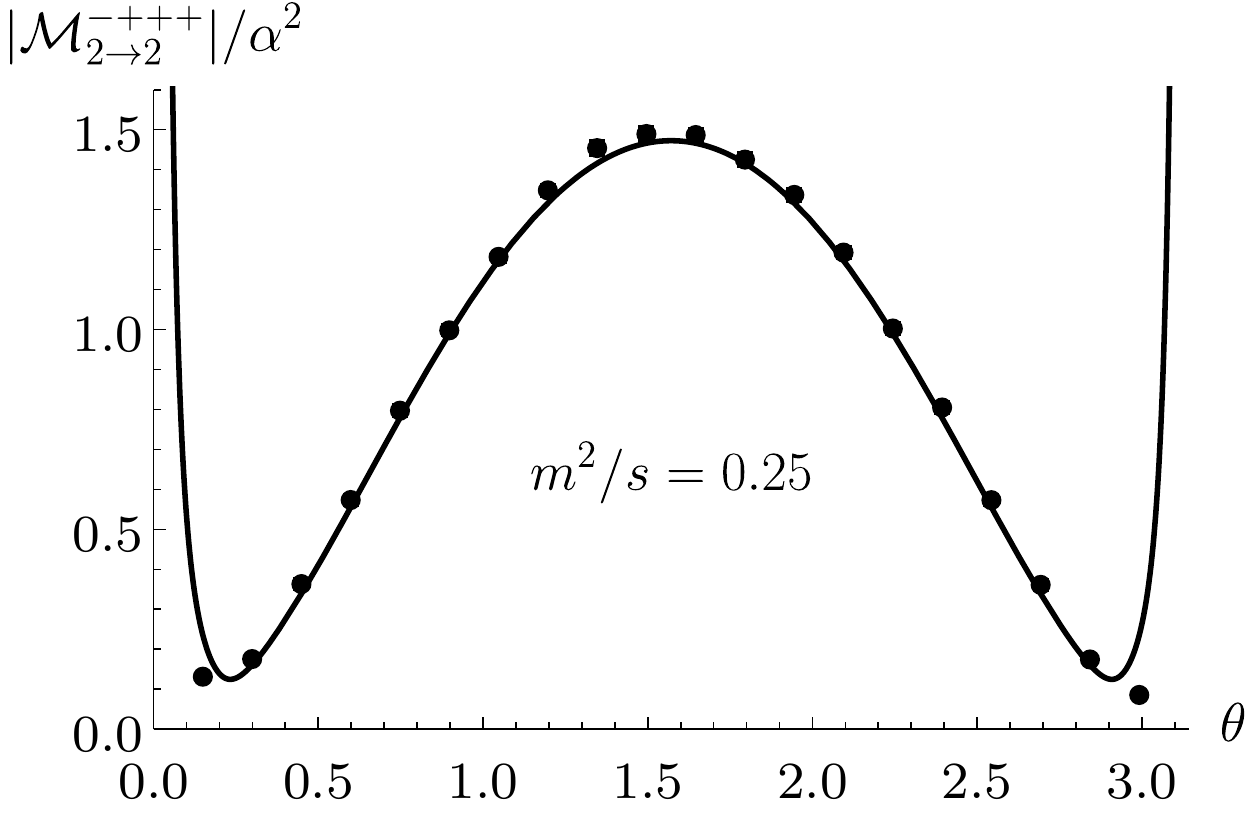}
}
\caption{\label{mppp} Four-photon amplitudes for $-+++$ helicities and different fermion masses. 
The curve is the analytic result, $\theta$ is the scattering angle and the points are the result of numerical integration and extrapolation.}
\end{figure}

\begin{figure}[!h]
\centering
\subfigure{
\includegraphics[width=.35\textwidth]{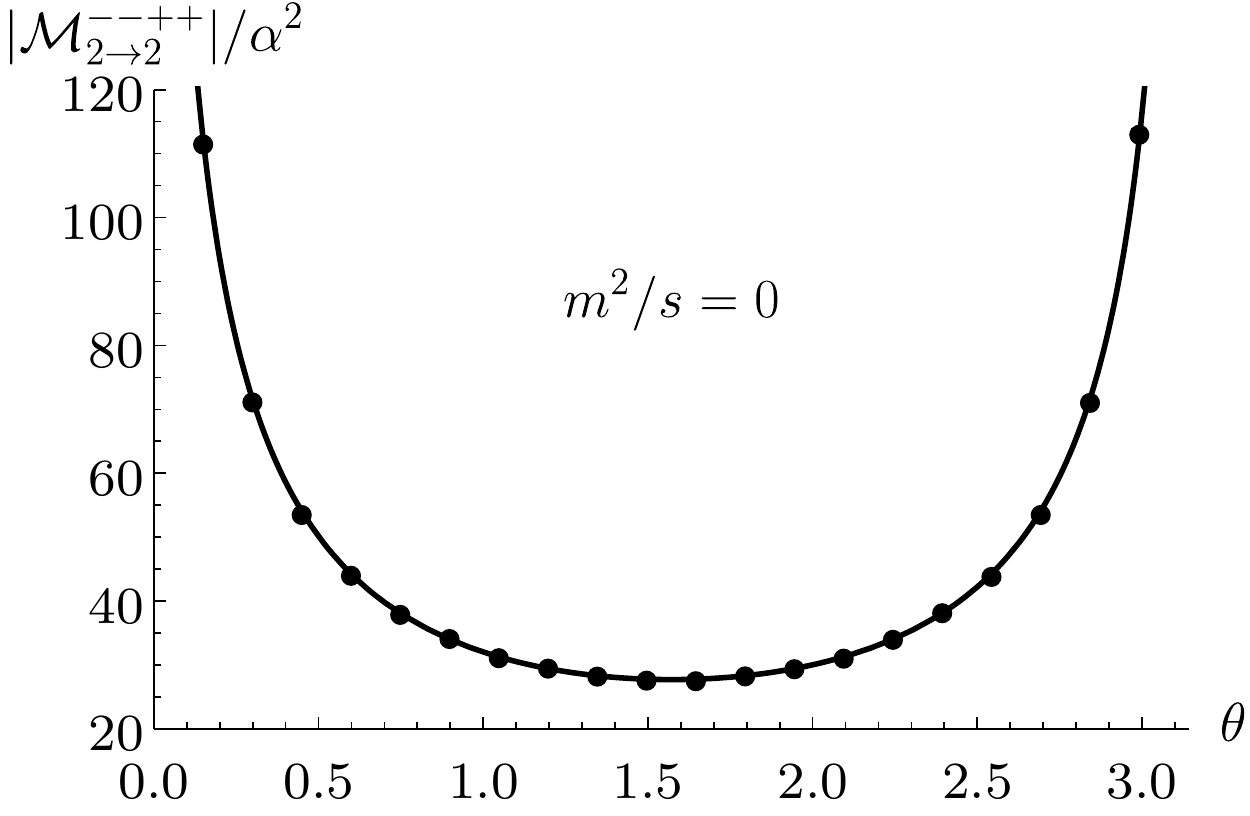}
}
\hspace{50pt}
\subfigure{
\includegraphics[width=.35\textwidth]{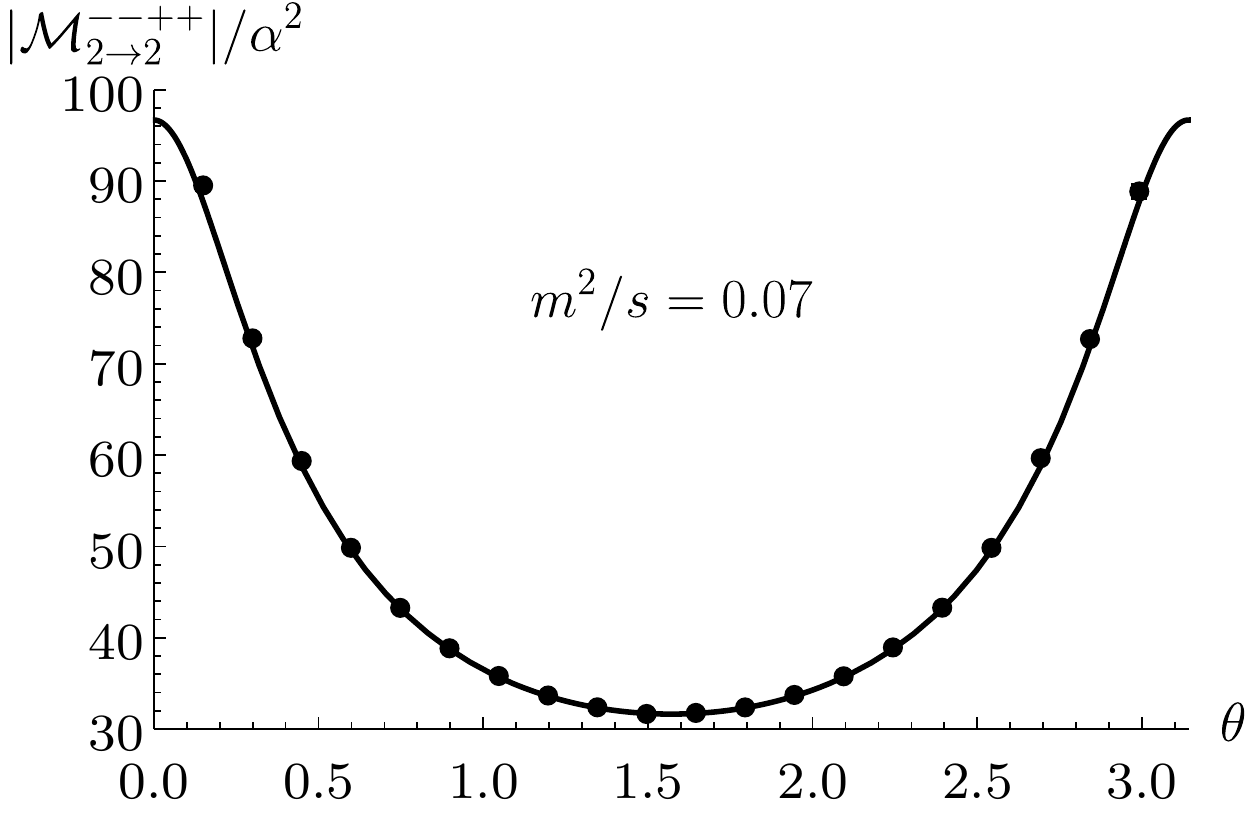}
} \\
\subfigure{
\includegraphics[width=.35\textwidth]{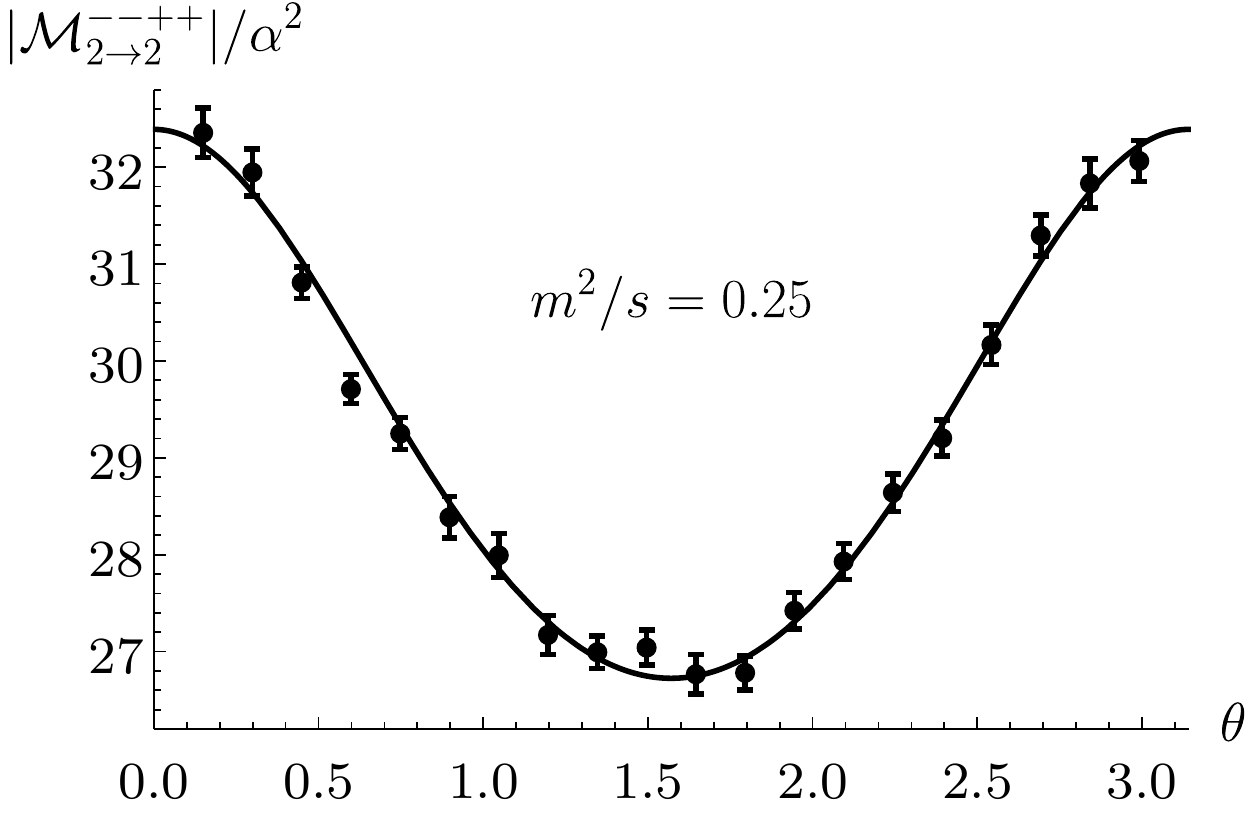}
}
\hspace{50pt}
\subfigure{
\includegraphics[width=.35\textwidth]{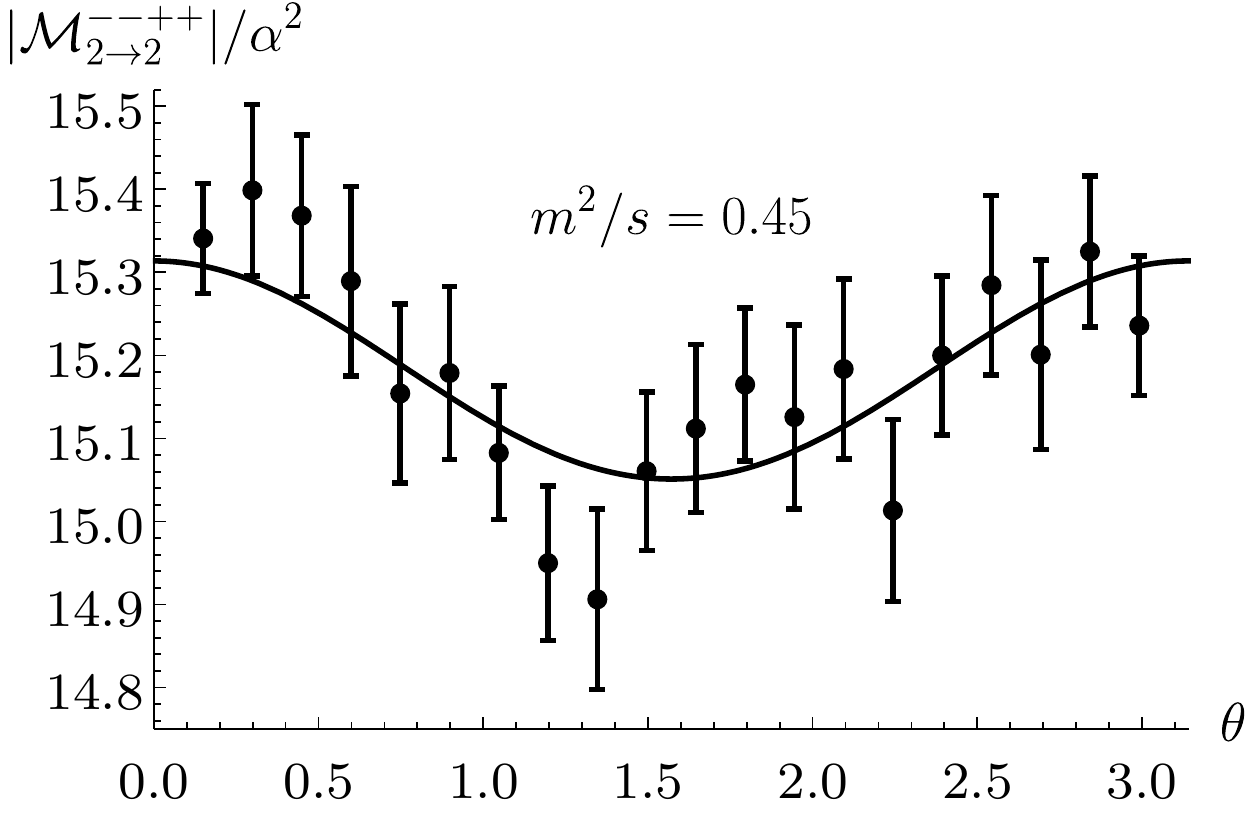}
}
\caption{\label{mmpp} Four-photon amplitudes for $--++$ helicities and different fermion masses. 
The curve is the analytic result, 
$\theta$ is the scattering angle
and the points are the result of numerical integration and extrapolation.}
\end{figure}

\begin{figure}[!h]
\centering
\subfigure{
\includegraphics[width=.4\textwidth]{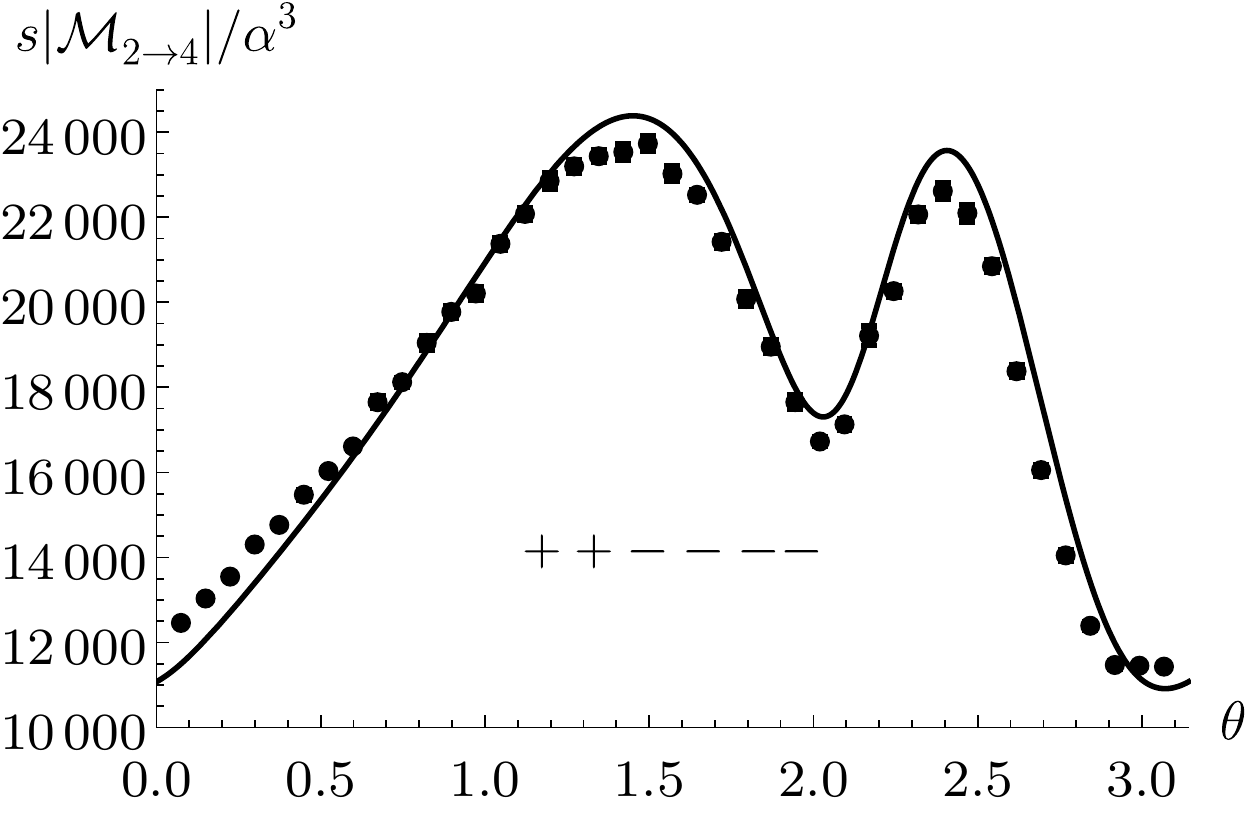}
}
\hfill
\subfigure{
\includegraphics[width=.4\textwidth]{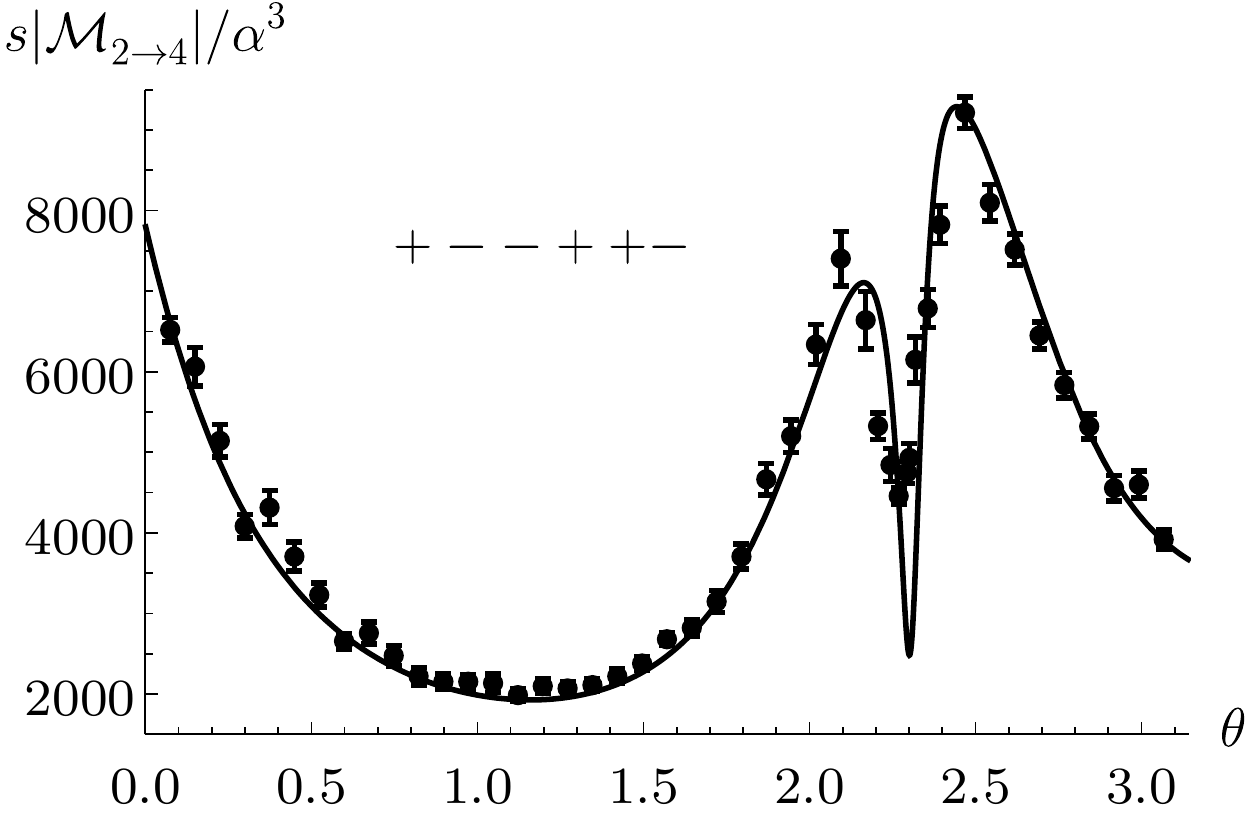}
}
\caption{\label{six} Six-photon amplitudes with massless fermions. 
An arbitrarily chosen final state from Eq.~(\ref{state}) was rotated
about the $y$ axis through the angle $\theta$.
The curve is the analytic result of Refs.~\cite{Mahlon1, Mahlon2} and \cite{Binoth} for $++----$ and
$+--++-$ helicities, respectively. The points are the result of numerical integration and extrapolation.}
\end{figure}

The results for $N=4$ and $N=6$ are given in Figs.~\ref{pppp}--\ref{six}.
As expected, for $N=5$ we got the result consistent with zero, i.e.~good cancellation between $\lim_{\epsilon \to 0} \mathcal{M}_\text{IR}(\epsilon)$ and $\mathcal{M}_\text{UV}$ parts.

\newpage

\begin{figure}[!h]
\centering
\subfigure{
\includegraphics[width=.4\textwidth]{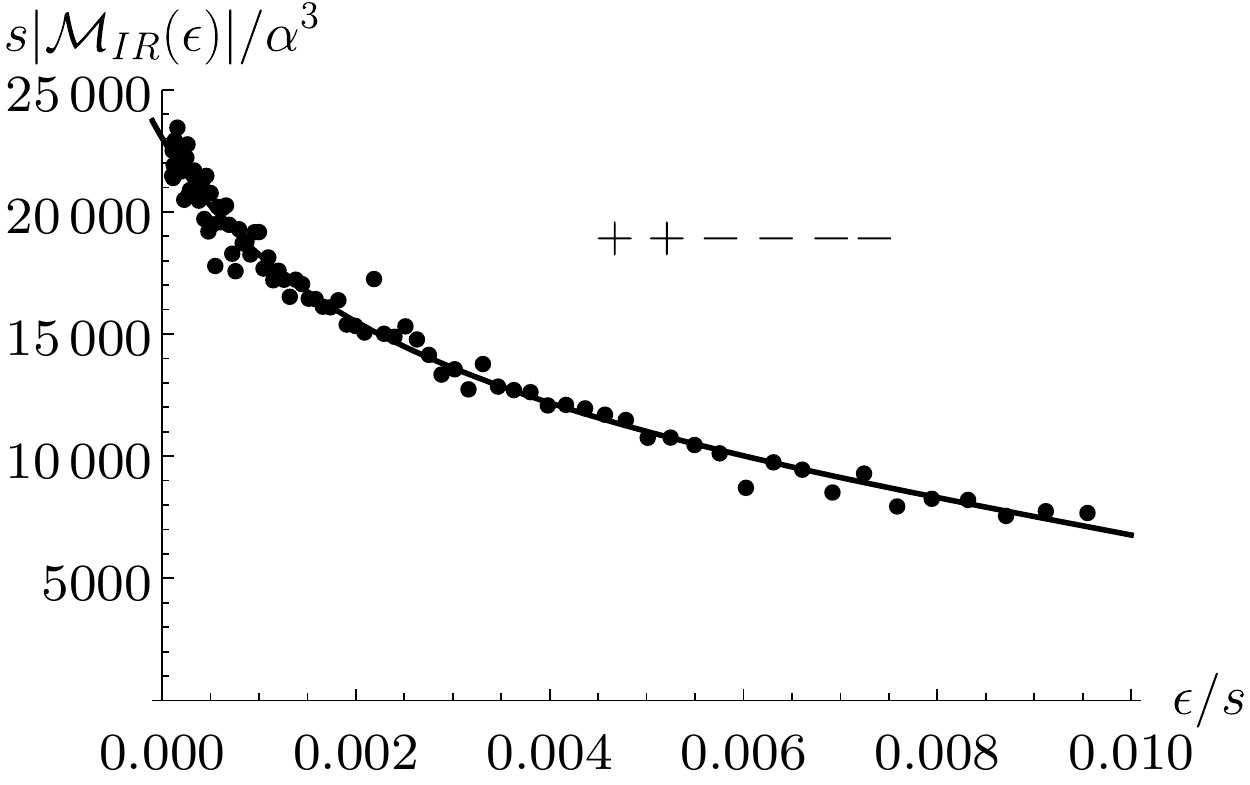}
}
\hfill
\subfigure{
\includegraphics[width=.4\textwidth]{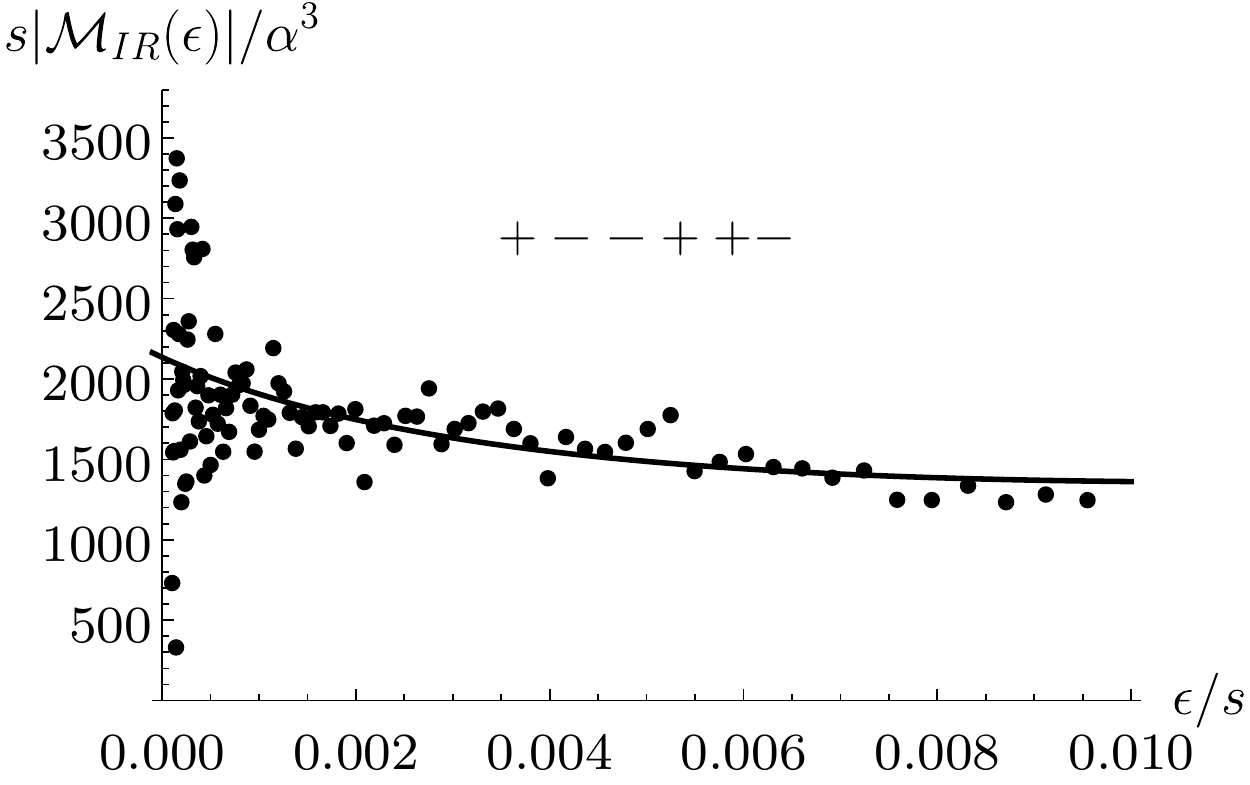}
}
\caption{\label{convergence} The IR part of the amplitude as function of $\epsilon$. These are two typical situations
for six-photon amplitudes. The points are the result of numerical integration given in the second term of Eq.~(\ref{4})
where for this case there are 720 terms in the integrand (120 diagrams with six residues each). The points were generated 
by using $10^6$ Monte Carlo points and precision of $10^{-2}$ for each $\epsilon \in [10^{-4}s,10^{-2}s]$, where $s$ is total energy squared. 
The curve is the Pad\'e approximant of order $[ 2/1 ]$ fitted to the numerical points.}
\end{figure}

Being very simple and straightforward, the method is highly acceptable. The fundamental concern about its practical use is the rate
of convergence of the sequence ${\cal M}_{\rm IR}(\epsilon_i)$, especially because we have to use not-so-small values of $\epsilon_i$
in order to accomplish stable numerical integration for the chosen number of Monte Carlo points. 
To get a sense of the rate of convergence (from our experience, it depends on the process, kinematics, etc.)
in Fig.~\ref{convergence} we present two typical situations for $N=6$. Despite nonambitious demands on
numerical integration precision ($10^{-2}$) and consequent large scattering of the points seen in Fig.~\ref{convergence}, the final results (Fig.~\ref{six}) are in good agreement with the analytic calculations. One wonders what can be accomplished if more integration points and more ambitious numerical integration is used. 

\subsection{Higgs decay}

As another example, we used our method to calculate the Higgs boson decay into two photon $H \to \gamma \gamma$ via $W$ boson loop. There has been some controversies concerning the amplitude of this process. It has originally been calculated in the Feynman--'t Hooft gauge ($R_\xi$ gauge with $\xi = 1$) \cite{Higgs1, Higgs2}. In this case, there are 26 different Feynman diagrams contributing to the amplitude (these diagrams include the unphysical particles such as ghosts and Goldstone bosons propagating in the loop) which is given by the expression
\begin{equation}
\mathcal{M}_{\xi = 1} = \frac{e^2 g}{(4\pi)^2 m_W} \left[(k_1 \cdot k_2)g_{\mu \nu} - k_{1\nu} k_{2 \mu} \right] \epsilon_1^{\mu} \epsilon_2^{\nu} \left\{ 2 + \frac{3}{\tau} + \frac{3}{\tau} \left(2 - \frac{1}{\tau} \right) \arcsin^2 (\sqrt{\tau}) \right\}, \label{7}
\end{equation}
where $g$ is the $HW$ coupling constant, $m_W$ the mass of the $W$-boson, $k_{1,2}$ and $\epsilon_{1,2}$ momenta and polarizations of the photons and $\tau = (m_H / (2 m_W))^2$, $m_H$ being the Higgs mass. Note that for $\tau > 1$, the $\arcsin$ function has to be analytically continued.

More recent calculations \cite{Higgs3, Higgs4}, using the unitary gauge ($R_\xi$ gauge with $\xi \to \infty$) claimed the correct amplitude for the Higgs decay process to be
\begin{equation}
\mathcal{M}_{\xi \to \infty} = \frac{e^2 g}{(4\pi)^2 m_W} \left[(k_1 \cdot k_2)g_{\mu \nu} - k_{1\nu} k_{2 \mu} \right] \epsilon_1^{\mu} \epsilon_2^{\nu} \left\{\frac{3}{\tau} + \frac{3}{\tau} \left(2 - \frac{1}{\tau} \right) \arcsin^2 (\sqrt{\tau}) \right\}, \label{8}
\end{equation}
in contradiction with Eq.~\eqref{7}. The advantage of using the unitary gauge is the absence of unphysical particles propagating in the loop, which reduces the number of different Feynman diagrams to 6.

\begin{figure}[!h]
\centering
\subfigure{
\includegraphics[width=.4\textwidth]{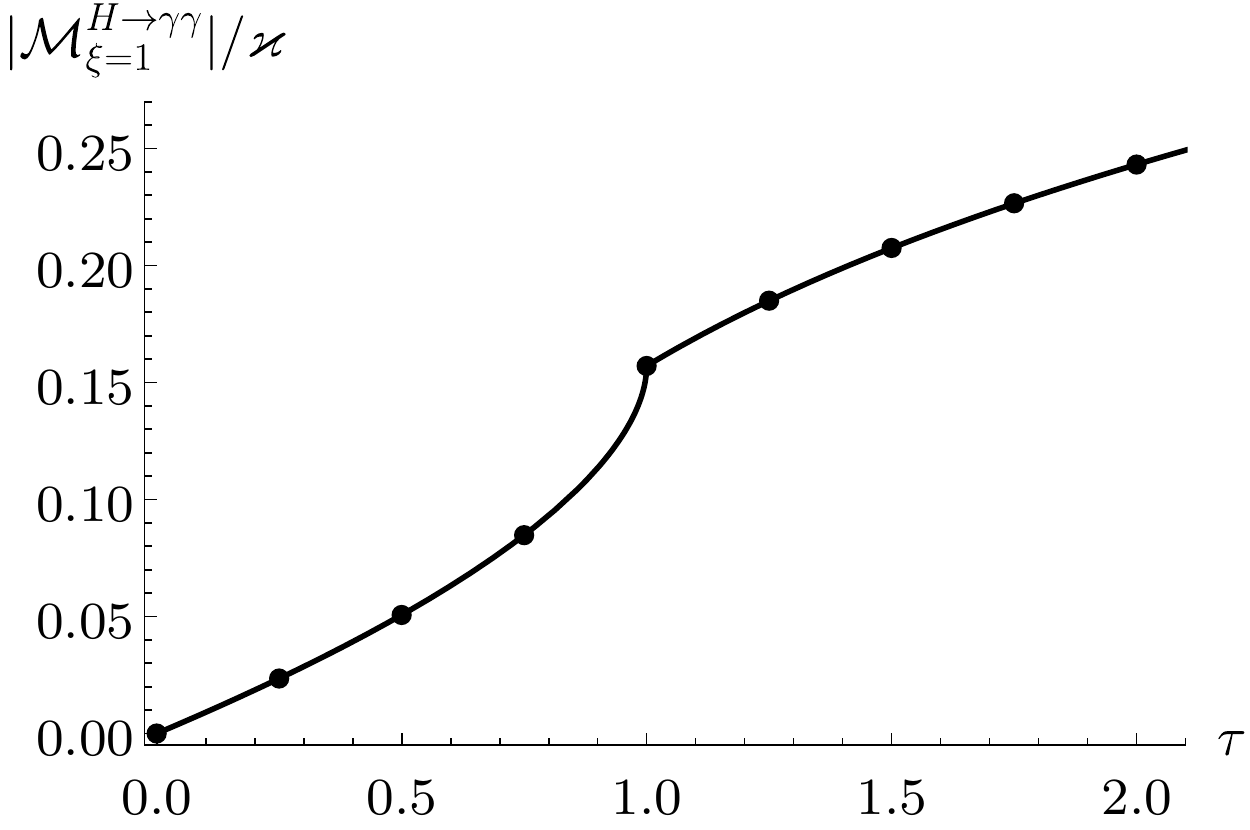}
}
\hfill
\subfigure{
\includegraphics[width=.4\textwidth]{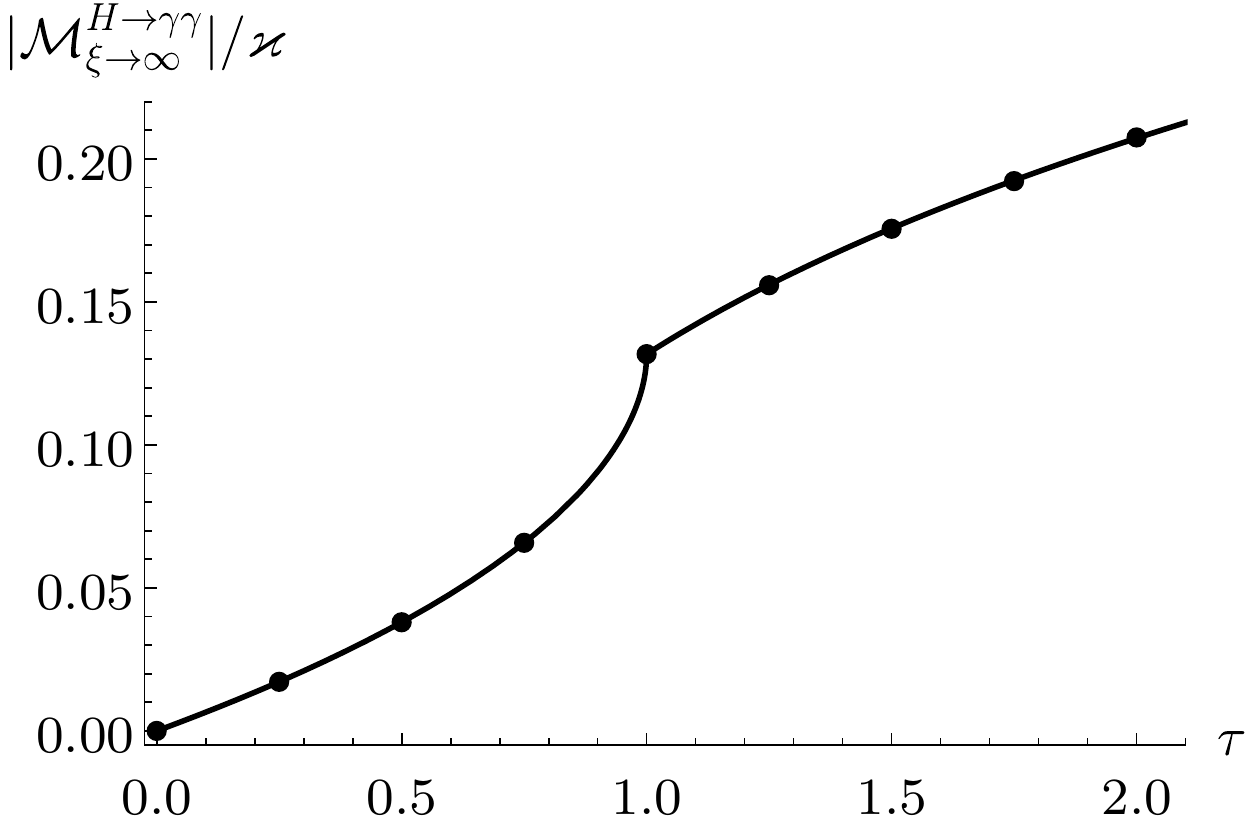}
}
\caption{\label{Higgs} Higgs decay amplitudes in both Feynman--'t Hooft ($\xi = 1$) and unitary ($\xi \to \infty$) gauges. The curves are analytical results given in Eqs.~\eqref{7} and \eqref{8}, while the points are obtained by numerical integration and extrapolation. The photons have the same helicity and we introduced the notation $\varkappa = e^2 g m_W$.}
\end{figure}

Leaving the question of the correct expression for the amplitude aside (we refer the reader to \cite{Higgs5} and references therein), we used our method to calculate the amplitude in both Feynman--'t Hooft and unitary gauges. Since the decay $1 \to 2$ is kinematically trivial, we calculated the amplitude as a function of the parameter $\tau$ (although it is now known that the physical value is $\tau = 0.6$). Our results are given in Fig.~\ref{Higgs}.

\section{Conclusions}
A new, completely numerical approach to calculating one-loop amplitudes has been presented. Using the Feynman $\rm i \epsilon$ term as a regulator of singularities and calculating the complete amplitude at once, we extrapolate the results to get the physical amplitude in the limit $\epsilon \to 0$. For such a low-level approach, we obtained a surprisingly good agreement with the analytical calculation, which we demonstrated on the benchmark process $2 \gamma \to (N-2) \gamma$, as well as on the Higgs decay amplitude $H \to \gamma \gamma$. In comparison to the existing direct numerical methods, our method is very simple and intuitive and works equally well for massless and massive cases. Also, we have shown that it is possible to obtain trustworthy results even with low requirements on integration reliability and precision. This suggests stability and robustness of the method. At the present level of development, direct numerical methods for calculation of the one-loop amplitudes are behind methods based on reduction to master scalar integrals. However, direct numerical approach only recently entered the scene and there is certainly room for improvement.

\section{Acknowledgements}
We are grateful to D.~Horvati\'c for useful discussions and providing the additional hardware for numerical calculations. The work
is  supported  by  the  Croatian  Science  Foundation  (HrZZ)  project ``Physics  of  Standard
Model  and  Beyond", Project No.~HrZZ 5169 and by the H2020 CSA Twinning project No.~692194, RBI-T-WINNING.

\bibliography{bibliopaper}

\end{document}